\documentclass[article]{paper}

\usepackage[utf8]{inputenc}
\usepackage{amsfonts}
\usepackage{bbm}
\usepackage{amsmath}
\usepackage{amssymb} 
\usepackage{mathtools}
\usepackage{hyperref}
\hypersetup{
	pdfmenubar=true,	    
	pdffitwindow=false,     
	pdfstartview={FitH},    
	colorlinks=true,        
	linkcolor=blue,         
	citecolor=red,          
	filecolor=blue,         
	urlcolor=blue      
}
\usepackage{cleveref}
\usepackage{graphicx}
\usepackage{rotating}
\usepackage{tikz}
  \usetikzlibrary{patterns}
  \usetikzlibrary{spy}
  \usetikzlibrary{backgrounds}
    \pgfdeclarelayer{foreground}
    \pgfsetlayers{background,main,foreground}
  \usetikzlibrary{decorations.markings}
  \usetikzlibrary{arrows}
\usepackage{pgfplots}
  \pgfplotsset{compat=newest}
\usepackage{lscape}
\usepackage[section]{placeins}
\usepackage{paralist}
\usepackage{booktabs}
\usepackage{makecell}
\usepackage{paperacronyms}
\usepackage{todonotes}
\usepackage{tablefootnote}

\newcommand{\stochastic}[1]{\mathbbm{#1}}
\newcommand{\Expect}{\operatorname{\mathbb{E}}}
\DeclarePairedDelimiterX{\infdivx}[2]
  {\delimsize[}{\delimsize]}{#1\;\delimsize\vert\;#2}
\newcommand{\cExpect}{\Expect\infdivx}
\newcommand{\estim}[1]{\widehat{#1}}
\newcommand{\project}[1]{\Bar{{#1}}}

\newcommand{\Real}{\mathbb{R}}

\newcommand{\RecSpace}{X}
\newcommand{\signal}{f}
\newcommand{\stsignal}{\stochastic{\signal}}
\newcommand{\DataSpace}{Y}
\newcommand{\data}{g}
\newcommand{\sinogram}{g}
\newcommand{\stdata}{\stochastic{\data}}
\newcommand{\datanoise}{e}
\newcommand{\stdatanoise}{\stochastic{\datanoise}}
\newcommand{\argmin}{\operatorname*{arg\,min}}
\newcommand{\oper}[1]{\operatorname{\mathcal{#1}}}
\newcommand{\ForwardOp}{\oper{A}}

\newcommand{\Loss}{\oper{L}}
\newcommand{\DataLoss}{\Loss_{\DataSpace}}
\newcommand{\RecLoss}{\Loss_{\RecSpace}}
\newcommand{\RecOp}{\oper{R}}
\newcommand{\RegFunc}{\oper{S}}

\newcommand{\ImageWidth}{I_w}
\newcommand{\ImageHeight}{I_h}
\newcommand{\NumSlices}{N_z}
\newcommand{\NumSourcePos}{N_\varphi}
\newcommand{\DetWidth}{D_w}
\newcommand{\DetHeight}{D_h}
\newcommand{\NumBlocks}{N_s}
\newcommand{\NumAngInLap}{N_a}
\newcommand{\Lap}{\data}
\newcommand{\Block}{\signal}
\newcommand{\ImageBlockThickness}{N_{t}}
\newcommand{\ProjOp}{\oper{P}}

\def\PrimalSpace{{X}}

\def\primal{{f}}
\def\dual{{u}}
\def\fluxCorr{{c}}

\def\PrimalMappingOriginal{{\Lambda}}
\def\DualMappingOriginal{{\Gamma}}
\def\Dim{{N}}
\def\PrimalDim{{\Dim_p}}
\def\DualDim{{\Dim_d}}

\def\NumSteps{M}
\def\nlaps{k}

\makeatletter
\providecommand{\maketitle}{}
\renewcommand{\maketitle}{%
	\par
	\begingroup
	\renewcommand{\thefootnote}{\fnsymbol{footnote}}
	\renewcommand{\@makefnmark}{\hbox to \z@{$^{\@thefnmark}$\hss}}
	\long\def\@makefntext##1{%
		\parindent 1em\noindent
		\hbox to 1.8em{\hss $\m@th ^{\@thefnmark}$}##1
	}
	\thispagestyle{empty}
	\@maketitle
	\@thanks
	\endgroup
	\let\maketitle\relax
	\let\thanks\relax
}

\providecommand{\@maketitle}{}
\renewcommand{\@maketitle}{%
	\vbox{%
		\hsize\textwidth
		\linewidth\hsize
		\vskip 0.1in
		\centering
		{\LARGE\bf \@title\par}
		\def\And{%
		\end{tabular}\hfil\linebreak[0]\hfil%
		\begin{tabular}[t]{c}\bf\rule{\z@}{24\p@}\ignorespaces%
		}
		\def\AND{%
		\end{tabular}\hfil\linebreak[4]\hfil%
		\begin{tabular}[t]{c}\bf\rule{\z@}{24\p@}\ignorespaces%
		}
		\begin{tabular}[t]{c}\bf\rule{\z@}{24\p@}\@author\end{tabular}%
		\vskip 0.3in \@minus 0.1in
	}
}
\makeatother

\usepackage{fancyhdr}

\fancypagestyle{FirstPage}{
\lfoot{\textbf{This work has been submitted to the IEEE for possible publication. Copyright may be transferred without notice, after which this version may no longer be accessible.}} 
}

\begin{document}

\title{3D helical \acs{CT} Reconstruction with a Memory Efficient \acl{LPD} Architecture}

\author{Jevgenija Rudzusika\thanks{Supported by Swedish Foundation of Strategic Research grant AM13-0049, grant from the VINNOVA Open Innovation Hub project 2015-06759 and by Philips Healthcare.}\\ Department of Mathematics \\ KTH Royal Institute of Technology\\ Sweden\\ \href{mailto:jevaks@kth.se}{\texttt{jevaks@kth.se}} 
\And Buda Baji\'{c} \thanks{Supported by Ministry of Education, Science and Technological Development of the Republic of Serbia, project ON174008.}\\ Faculty of Technical Sciences \\ University of Novi Sad\\ Serbia\\ \href{mailto:buda.bajic@uns.ac.rs}{\texttt{buda.bajic@uns.ac.rs}} \\
\And Thomas Koehler \\ Philips Research \\ Germany \\
\href{mailto:thomas.koehler@philips.com}{\texttt{thomas.koehler@philips.com}} \\
\And Ozan \"Oktem \\ Department of Mathematics \\ KTH Royal Institute of Technology \\ Sweden \\ \href{mailto:ozan@kth.se}{\texttt{ozan@kth.se}}  }

\maketitle

\acresetall
\begin{abstract}
    Deep learning based computed tomography (CT) reconstruction has demonstrated outstanding performance on simulated 2D low-dose CT data. This applies in particular to domain adapted neural networks, which incorporate a handcrafted physics model for CT imaging. Empirical evidence shows that employing such architectures reduces the demand for training data and improves upon generalisation. However, their training requires large computational resources that quickly become prohibitive in 3D helical CT, which is the most common acquisition geometry used for medical imaging. Furthermore, clinical data also comes with other challenges not accounted for in simulations, like errors in flux measurement, resolution mismatch and, most importantly, the absence of the real ground truth. The necessity to have a computationally feasible training combined with the need to address these issues has made it difficult to evaluate deep learning based reconstruction on clinical 3D helical CT. This paper modifies a domain adapted neural network architecture, the Learned Primal-Dual (LPD), so that it can be trained and applied to reconstruction in this setting. We achieve this by splitting the helical trajectory into sections and applying the unrolled LPD iterations to those sections sequentially. To the best of our knowledge, this work is the first to apply an unrolled deep learning architecture for reconstruction on full-sized clinical data, like those in the Low dose CT image and projection data set (LDCT). Moreover, training and testing is done on a single GPU card with 24GB of memory.
\end{abstract}
\acresetall

\section{Introduction}
\thispagestyle{FirstPage}
Clinical \ac{CT} aims to computationally recover an image representing the interior anatomy of a subject by taking X-ray projections from different directions.

Early approaches were inherently 2D in the sense that they aimed to recover a specific 2D cross section of the subject.  
To get a 3D image required one to repeat this for multiple adjacent 2D cross sections (step and shoot method).
A major advance for 3D imaging came in the late 1980s with the introduction of helical scanning \cite{LaRivierea:2021aa}. 
Here, the subject moves continually through the gantry while the x-ray source and corresponding detector is rotating.
This corresponds to having an x-ray source on a helical path around the subject. 
Compared to step and shoot acquisition, helical \ac{CT} scanning is much faster, e.g.\@ one can scan many organs in a single breath-hold, thus reducing pulmonary motion artefacts. Helical scanning is therefore the de facto standard acquisition method for clinical \ac{CT}.

It is challenging to design reconstruction methods that are capable of taking full advantage of the data acquired in 3D helical \ac{CT} \emph{without} compromising on computational feasibility. 
This is mostly due to the combination of having a complicated acquisition geometry and the large scale nature of the problem, the arrays for storing data and 3D image can in clinical 3D helical \ac{CT} easily reach $10^8$--$10^9$ elements.

Theoretically exact analytic reconstruction methods have fast reconstruction run-time, but they fail to make use of \emph{all} the measured data without resorting to approximations.
In contrast, model based iterative or variational approaches can use all of the data, but they have too long reconstruction run-time. 
Deep learning based methods form a third paradigm for reconstruction that outperform model based approaches regarding the image quality while having a reconstruction run-time that is closer to the analytic methods \cite{Arridge:2019aa,Willemink:2019aa,LaRivierea:2021aa}.
Nevertheless, the substantial demand for computational resources during training  constrains 
the size of problems one can address without using \ac{GPU} super-computing resources and extensive software engineering.

The overall aim of this paper is to develop a deep learning based reconstruction method that \begin{inparaenum}[(a)] \item delivers high image quality, \item can be trained on \emph{high-end consumer \ac{GPU} hardware}, \item and can be applied to the clinical 3D helical \ac{CT} data. \end{inparaenum} We start out from the \ac{LPD} architecture \cite{Adler:2018aa} that has shown state-of-the-art performance in 2D \ac{CT} reconstruction.  Our \emph{main contributions} are the following:   
\begin{itemize}
    \item We show how to overcome technical challenges and scale up the method to be applicable to 3D helical CT. This includes finding the right balance between \ac{GPU} memory and computational time required for training;
    \item We propose changes in the neural network architecture that are specifically suited for helical geometry and lead to additional performance gains; 
    \item We apply the method to clinical data from the \ac{LDCT} data set. This includes bridging the gap between the real data and simulations, while dealing with the absence of the actual ground truth.
\end{itemize}

\subsection{Computational feasibility}

The \ac{LPD} architecture is formed by truncating the iterative scheme given by the \ac{PDHG} optimization method, then replacing proximal operators with neural networks. The resulting deep neural network consists of blocks representing unrolled iterations. As with all deep neural networks, the drawback of \ac{LPD} is its large \ac{GPU} memory footprint during back-propagation, which is too large for end-to-end training on high-dimensional input data.
Its possible to reduce the \ac{GPU} memory footprint of a neural network in training at the cost of increased computational training time. The challenge is to keep the computational time within the practical limits. To exemplify, the training of our final method on a single GeForce RTX~3090 \ac{GPU} with 24GB of memory required 6 weeks! Omitting any of the steps outlined below renders unfeasible training time.

\paragraph{Check-pointing} We start with gradient check-pointing \cite{chen2016training}, i.e. saving activation values of some hidden layers during the forward pass of the network and recomputing the others during the back-propagation. However, moving the checkpoints to the \ac{CPU} memory leads to communication overhead, so saving layers that have many channels is inefficient, not to mention that the \ac{CPU} memory also has limits. Therefore, we only save inputs and outputs of each unrolled iteration as checkpoints.

\paragraph{Partitioning the data}
The dimension of clinical tomographic data is so large that even a single unrolled \ac{LPD} iteration does not fit into the memory of a high-end consumer \ac{GPU}. Therefore, we partition the projection data into sections and identify the sub-volumes in the image domain that correspond to those sections. Next, we apply unrolled \ac{LPD} iterations to these sub-volumes using only one data section at a time. As shown in \cref{subsec:ablation}, this modification of the original \ac{LPD} architecture is beneficial in the context of helical \ac{CT}.

\paragraph{Training on smaller samples} Even though the above steps are sufficient to process clinical projection data of any size, the training on full-sized data would take months, if not years. In order to facilitate faster training, we train against smaller samples of the projection data and corresponding 3D reconstructions. We investigate how the size of the training samples influence the performance and find that the optimal size of the projection data corresponds to two helical turns. The trained network generalizes well to much longer helical trajectories (12.5 - 36 turns in our test cases). 
Furthermore, we explore an alternative strategy, which leads to significant improvements: We apply the network using the same input size as it was trained for in a ``sliding window'' manner. The obtained partial reconstructions are then combined using weighted averaging as described in \cref{sec:TrainTest}.

\subsection{Clinical data}
The main challenge in supervised training of deep learning--based reconstruction methods against real clinical \ac{CT} data is the absence of the ground truth. We address this problem by training the method on simulated data and then evaluating it on the real data. We make the simulations more accurate than it is typically done by using the real acquisition geometries and noise statistics as in the \ac{LDCT} data set. In particular, this means that the method is trained to implicitly handle varying pitch and data signal-to-noise ratio that occur in the clinical setting. Secondly, we  apply corrections to the real data to address the remaining issues, such as, errors in flux measurement, objects outside the scans field of view and resolution mismatch. Finally, we discuss an alternative approach, i.e. using a full dose reference reconstruction as a proxy for the ground truth, in section \cref{sec:discussion}.

\section{Related work}

First reconstruction methods for helical 3D \ac{CT} were \emph{analytic methods} that combine back-projection with a convolution using a suitable (reconstruction) filter.
Since the filter in such a method depends on the acquisition geometry, early approaches relied on approximate recovery that combines re-binning to a simplified geometry, e.g., slice-wise 2D \cite{noo1999single} with 2D-\ac{FBP} or 3D-\ac{FDK} type of analytic reconstruction (see surveys \cite{Bontus:2009aa,Hsieh:2013aa} and \cite[Table~1.2]{Turbell:2001aa}).
Novel reconstruction filters for \ac{FBP} that provide theoretically exact recovery from helical \ac{CT} data where introduced in 2003 \cite{Katsevich:2003aa,Katsevich:2006aa}.

A fundamentally different approach is to define  reconstruction as a solution to an optimization problem that involves minimizing data discrepancy and accounts for prior assumptions through additional regularizing components. These \emph{iterative model based methods} have become increasingly popular since 2015 and are nowadays preferred over analytic reconstruction methods in clinical \ac{CT} \cite{Mileto:2019aa}.
Here the acquisition geometry is explicitly encoded in the forward/back-projection model. However,
arrays needed to store projection and image data in helical \ac{CT} are very large, so the key challenge is to handle the computational burden.
Currently, most iterative model based methods are stopped long before iterates have converged.
The extensive mathematical theory that offers convergence guarantees and error estimates for many of these model based methods \cite{Scherzer:2009aa} is therefore not applicable in this setting.

Recently there is an increasing interest in data driven deep learning based approaches for \ac{CT} reconstruction. Much of the development is catalysed by the possibility to significantly improve image quality in low-dose setting \emph{without} compromising on computational feasibility. However, these methods require extensive computational resources for adjusting free parameters during training. Given the size of the input data in helical \ac{CT}, limited \ac{GPU} memory is a major problem.

Certain features in an architecture of a neural network allow to reduce its memory footprint at the cost of additional computations.
As an example, invertibility is a property that allows  to recompute activations of intermediate layers, while performing back-propagation \cite{gomez2017reversible}. Inspired by this idea, \cite{rudzusika2021invertible} modified the architecture of state-of-the art \ac{LPD} method \cite{Adler:2018aa}, so that its unrolled iterations become invertible. As a result, the memory footprint of the method was reduced to the footprint of one invertible block. \cite{moriakov2023end} used invertibility combined with patch-wise evaluation of convolutional operations within the invertible blocks, which allowed to shift the balance between memory and computational time even further. The advantage of using invertible networks over  gradient check-pointing is that there is no need for additional \ac{CPU}-\ac{GPU} communication. The drawback is that the modification likely results in reduced model capacity so its performance is expected to decrease when compared to the original \ac{LPD} architecture \cite{rudzusika2021invertible}. In addition, each invertible block has to be executed three times, while each network block between the checkpoints has to be recomputed only twice.
Another approach proposed in \cite{thies2022learned} was inspired by neural \acp{ODE} \cite{chen2018neural}. Here, a solution to a reconstruction problem is represented by a function that evolves over time and the derivative of this function is represented by a neural network. The final solution can be obtained through numerical integration, which results in repeated applications of the network with additive skip-connections. This architecture does not require storing activations of intermediate layers as the gradients can be computed by an \ac{ODE} solver instead of traditional back-propagation. In this case, the memory footprint of the method is approximately equal to the footprint of the network representing the derivative. So far, all the methods described above have been applied in 3D \ac{CT} reconstruction with a circular acquisition geometry. However, their application to helical geometry requires additional improvements.   

Another approach to reduce the memory footprint is to reduce the amount of input data that is processed by the neural network at the same time. A natural approaches for achieving this is splitting the input data into parts and down-sampling. The down-sampling strategy has been explored in \cite{hauptmann2020multi}, where \acl{LGD} is applied in different scales, starting from a coarse reconstruction and then gradually increasing resolution to reconstruct small details. However, since the proposed architecture involves a block that acts on the full resolution, the memory footprint of this method is still large.

Splitting the tomographic data is not a new concept in tomography. In \acl{OS} methods  \cite{Hudson:1994aa, erdogan1999ordered, kim2014combining, ahn2006convergent} only a section of the data is used to update an image volume during one iteration of iterative reconstruction. This leads to initial acceleration of various optimization methods. Recently, this idea has been used to modify the \ac{LPD} architecture \cite{tang2021stochastic} with the goal of improving computational efficiency of the method. However, \ac{LSPD} method presented in \cite{tang2021stochastic}, relies on a stochastic choice of the section of the data, while in this work we select the sections subsequently in order to ensure that all the data has been used (which might not be the case in a stochastic setting, given relatively few unrolled iterations). 

The idea of splitting image volumes is partially explored in \cite{wu2019computationally}, where images are split into parts (patches) and those are used within greedy training of \acl{LGD}. However, the greedy training often leads to sub-optimal performance compared to an end-to-end approach, as it was also observed \cite{rudzusika2021invertible}. 
Furthermore, an approach of splitting image volumes for parallelization purposes is presented in \cite{murthy2020block}. Here, an image volume is split into non-overlapping blocks, so that only neighboring blocks share some tomographic data. This is used to split computations across multiple \ac{GPU}s with reduced communication overhead. In contrast, our work in this paper focuses on performing computations on one \ac{GPU}.

A computationally efficient way to use deep learning for image reconstruction in 3D is to augment an analytical reconstruction method, such as \ac{FBP}, \cite{lagerwerf2020computationally, wagner2022benefit, kosomaa2022projection}. This can be done by learning the convolutional kernel for filtering the data and/or by applying the neural networks for pre-processing the data and post-processing the reconstruction.
A lot of deep learning methods address only the post-processing step by learning the mapping between low-dose and normal-dose reconstructions obtained with analytical methods. In particular, very few works have utilized the clinical projection data in the \ac{LDCT} data set \cite{mccollough2020}, even fewer without re-binning to a parallel 2D geometry. One exception is \cite{kosomaa2022projection}, they trained an \ac{FBP}-based architecture using an unsupervised loss in the data domain. As a result the training required neither ground truth reconstructions, nor a reference reconstruction method. 
As we discuss in \cref{sec:discussion}, the absence of the ground truth reconstructions is indeed an important issue to address. We find the unsupervised training strategies as in \cite{kosomaa2022projection} to be very promising and note that it can be applied to the \ac{LPD} as well. However, we leave it out of the scope of this paper.

\section{Method}\label{sec:method}

First, we provide the theoretical foundations for deep learning based reconstruction methods. Then we outline the original \ac{LPD} architecture and describe the proposed modifications. Lastly, we propose different strategies for training and inference using helical \ac{CT} data.

\subsection{Theoretical foundations}\label{sec:TheoFound}
Learned iterative reconstruction uses principles from statistical decision theory to learn an `optimal' reconstruction method from training data \cite[Section~5]{Arridge:2019aa}.
Consider a setting where one has access to supervised training data in the form of i.i.d.\@ samples $(\signal_1,\data_1), \ldots, (\signal_n,\data_n) \in \RecSpace \times \DataSpace$ of $(\stsignal,\stdata)$ where $\stsignal$ and $\stdata$ are $\RecSpace$- and $\DataSpace$-valued random variable generating 3D images and helical \ac{CT} data, respectively, and    
\begin{equation}\label{eq:Forward_model}
\stdata = \ForwardOp(\stsignal) + \stdatanoise.
\end{equation}
The mapping $\ForwardOp \colon \RecSpace \to \DataSpace$ (ray transform) models how a 3D image gives rise to helical \ac{CT} data in absence of observation errors, and the random variable $\stdatanoise$ represents observation error.

Our aim is now to learn the `best' reconstruction method from the above supervised training data. 
To proceed, one has a pre-defined parametrised family $\{ \RecOp_{\theta} \}_{\theta}$ of possible reconstruction methods $\RecOp_{\theta} \colon \DataSpace \to \RecSpace$.
The `best' one can be taken as a Bayes' estimator, which for image-loss $\RecLoss \colon \RecSpace \times \RecSpace \to \Real$ is defined as
\begin{equation}\label{eq:BayesEst}
\RecOp_{\estim{\theta}} \colon \DataSpace \to \RecSpace
\quad\text{where}\quad
\estim{\theta} \in \argmin_{\theta} 
  \sum_{i=1}^n
    \RecLoss\bigl( \RecOp_{\theta}(\data_i), \signal_i \bigr).
\end{equation}

The choice of the image-loss $\RecLoss \colon \RecSpace \times \RecSpace \to \Real$ governs the type of estimator one seeks to approximate.
Choosing squared $\ell_2$-loss $\RecLoss(\signal,\signal'):= \Vert \signal-\signal' \Vert_2^2$ means that the learned reconstruction method approximates the posterior mean, i.e., 
\[ \RecOp_{\estim{\theta}}(\data) 
\approx \cExpect[\bigl]{\stsignal}{\stdata=\data}
\quad\text{where $\estim{\theta}$ is given by \cref{eq:BayesEst}.}
\]

A key component is the choice of parametrisation for the family of reconstruction methods $\RecOp_{\theta} \colon  \DataSpace \to \RecSpace$. 
One could select a generic \ac{DNN} architecture, consisting of fully connected layers, but if applied to helical \ac{CT}, such architectures easily become very large due to the size of the input and output data.
Training such \acp{DNN} would require vast amounts of data and the resulting learned reconstruction method, would most likely generalise poorly.
An alternative approach is to adapt the \ac{DNN} architecture for $\RecOp_{\theta}$.
A basic way to do this is to consider $\RecOp_{\theta} :=  \oper{C}_{\theta} \circ \ForwardOp^{\dagger}$  where $\ForwardOp^{\dagger} \colon \DataSpace \to \RecSpace$ is a hand-crafted approximate inverse of $\ForwardOp$ and $\oper{C}_{\theta} \colon \RecSpace \to \RecSpace$ is some 3D image-to-image post-processing operator that is learned from training data.
Popular architectures in imaging for the latter are based on \acp{CNN}, like U-Net \cite{jin2017deep}.

A different path to domain adaptation is through \emph{unrolling}. 
The idea is to start with some iterative scheme, like one designed to minimize $\signal \mapsto \bigl\Vert \ForwardOp(\signal) - \data\bigr\Vert_2^2$.
The next step is to truncate this scheme and replace the handcrafted updates with possibly shallow \acp{CNN} (unrolling).
$\RecOp_{\theta}$ is then a \ac{DNN} that is formed by stacking these shallow \acp{CNN} and coupling them with  $\ForwardOp$ and its adjoint $\ForwardOp^*$, which are explicitly given \cite[Sec.~4.9.1]{Arridge:2019aa}.
An example of this is the \ac{LPD} method \cite{Adler:2018aa} and variants thereof that are described in \cref{sec:LPD}.
Finally, note that \emph{unrolling is merely a way to select a \ac{DNN} architecture for $\RecOp_{\theta}$}.
In particular, training as in eq.~\eqref{eq:BayesEst} will \emph{not} yield a solution operator for an optimization problem. 

\subsection{Architectures}\label{sec:LPD}

The state-of-the-art unrolling scheme for \ac{CT} reconstruction is the \ac{LPD} network introduced in \cite{Adler:2018aa}. In the following subsections, we briefly describe the original method, \ac{LPD}, and technical steps that are sufficient to use this method in a 3D setting essentially without changing the architecture. Further we propose an improvement of the method specifically suited for helical geometry, \ac{LPDh}.

\subsubsection{The original \acl{LPD} architecture}

The \ac{LPD} architecture is inspired by the iterative scheme in the \ac{PDHG} algorithm \cite{chambolle2011first}. This architecture incorporates a forward operator and its adjoint into a \ac{DNN} by unrolling \cite{monga2021algorithm} a proximal primal-dual optimization scheme and replacing the proximal operators with \acp{CNN}. Hence, the \ac{LPD} architecture is domain adapted in the sense that the forward/back-projection operators encode an explicit physics based model for \ac{CT} data acquisition.
More precisely, the \ac{LPD} architecture is given in Algorithm~1, where $\NumSteps$ is the number of unrolling iterates, $\Dim$ is the total number of memory channels and superscripts (1) and (2) denote the 1st and the 2nd channel of the assigned variables. The functions $\DualMappingOriginal^i$ and $\PrimalMappingOriginal^i$ are \acp{CNN} that operate on the dual and the primal space, respectively. For each unrolled iteration $i$ these networks have the same architecture but different learned parameters. To simplify notation, we suppress the explicit dependence of $\DualMappingOriginal^i$ and $\PrimalMappingOriginal^i$ on the learned parameters ${\theta \in \Theta}$.

\begin{table}[t]
\centering
	\small
	\begin{tabular}[t]{@{\;}l}
	\hline
	\textbf{Algorithm~1}: \Acs{LPD} \cite{Adler:2018aa} 
	\tablefootnote{Note that this \Acs{LPD} formulation slightly differs from the one given in Alg.~3  in \cite{Adler:2018aa}, but it still reflects the actual implementation used in experiments in \cite{Adler:2018aa}.} 
	\\
	\hline
	1: Choose initial primal and dual variables \\
	\quad
        $(\primal_0, \dual_0)=\texttt{init}(\sinogram)$, where $(\primal_0, \dual_0) \in (\PrimalSpace^\PrimalDim, \DataSpace^\DualDim)$ \\
	2: \textbf{For} $i=1,2,\dots,\NumSteps$ \textbf{do}: \\
    3: \hspace{0.3cm} Dual update: $\dual_{i} = \dual_{i-1} + \DualMappingOriginal^i \left(\dual_{i-1}, \ForwardOp \primal_{i-1}^{(2)}, y\right)$ \\
	4: \hspace{0.3cm} Primal update: $\primal_{i} = \primal_{i-1} +  \PrimalMappingOriginal^i \left(\primal_{i-1}, \ForwardOp^\ast \dual_i^{(1)}\right)$ \\
	5: \textbf{return} $\primal_M^{(1)}$ \\
 [1mm]
    \hline
	\end{tabular}
\end{table}

In the original \Acs{LPD} paper \cite{Adler:2018aa}, $\DualMappingOriginal^i$ and $\PrimalMappingOriginal^i$ are convolutional neural networks with two hidden layers and parametric \ac{ReLU} activation functions. Here, we substitute the 2D convolutional layers by 3D convolutional layers and use simple \ac{ReLU} activation functions, because of the possibility to evaluate the function in-place without using additional memory.

Primal and dual variables $(\primal_0, \dual_0)$ are initialized with zeros.

\paragraph{Check-pointing} 
Gradient check-pointing is an ``off-the-shelf'' technique to reduce the memory footprint of a neural network during the back-propagation at the cost of computational time \cite{chen2016training}. Instead of storing activation values of the whole computational graph in memory, it is possible to save values of selected layers (checkpoints) and recompute the graph part by part from these layers. In a typical scenario, a part of the graph that has to be recomputed should fit into the \ac{GPU} memory. Therefore, when the memory is limited there is an incentive to split the graph into smaller parts. Furthermore, the checkpoints can be stored in \ac{CPU} memory. Unfortunately, copying the data between devices also takes time. Therefore, it is not beneficial to copy very large arrays. In the case of \ac{LPD}, certain balance is achieved when the checkpoints are done between the unrolled iterations, meaning that only inputs and outputs of the neural networks $\DualMappingOriginal^i$ and $\PrimalMappingOriginal^i$ are saved, while all intermediate computations are recomputed. The reason is that hidden layers inside the networks have much more channels than primal and dual variables $(\primal_{i}, \dual_{i}), i =1, \dots, M$ (typically 32 channels in the hidden layers and 5 channels for the primal and dual variables).

\paragraph{Splitting the layers} 
Despite a very large reduction in memory requirements achieved with check-pointing, it is still not sufficient for training \ac{LPD} on helical \ac{CT} data on single \ac{GPU} of standard capacity of 24 GB.
The reason is that the size of a typical helical scan of an abdomen is about 7 GB, while the size of a typical reconstruction is about 200 MB. In \ac{LPD} one check-pointed block ($\DualMappingOriginal^i$ or $\PrimalMappingOriginal^i$) is a shallow neural network applied either in data domain or in image domain. If this network has 2 hidden layers with $n$ channels, then the memory footprint will be at least $2n + 1$ multiplied by the size of an input (7 GB). Evidently, with such a large amount of input data, even a very shallow network does not fit into the memory.
One solution is to de-parallelize a \ac{CNN} by splitting the input arrays into spatial blocks and applying the network to these blocks sequentially \cite{moriakov2023end}. To ensure that the output values are not affected by the splitting, it is necessary to use a small overlap between blocks. The size of this overlap depends on the field of view of the output neurons and, in case of a 3 layer \ac{CNN} without strides, it is necessary to add 3 cells to each block from each side.

\subsubsection{\Acl{LPD} architecture for helical \ac{CT}} \label{subsec:LPDhos}

Here, we propose a different approach for ``de-parallelization'' of a single unrolled \ac{LPD} iteration. We partition the projection data into non-overlapping sections and identify the sub-volumes in the image domain that correspond to those sections. Inspired from the \ac{OS} optimization methods \cite{Hudson:1994aa}, we subsequently apply unrolled \ac{LPD} iterations to these sub-volumes using only one data section at a time. 
An advantage of this approach from the optimization perspective is that the updated part of the image serves as input to the next sub-iteration. In contrast, the original \ac{LPD} uses all the projection data and then simultaneously updates all voxels in the image domain. Looking from the deep learning perspective at this approach, we can say that the connectivity between spatially distant neurons within the neural network is increased without increasing the number of network parameters. Further follows a detailed description of the method.

Each element in the tomographic data corresponds to the attenuation of a ray propagating through an object. Mathematically (before discretization and without noise) this can be seen as an integral over a line $l$ defined in the image domain:
\begin{equation}
    \data(l) = \ForwardOp(\signal)(l)  = \int_{l} \signal\, ds.
\end{equation}
We process only a subset of these line integrals at a time that correspond to a contiguous section of data. These lines pass through the object of interest in a limited region in the image domain. Thus, instead of applying a neural network $\PrimalMappingOriginal^i$ to the whole image volume, we apply it only to a sub-volume, which is crossed by the aforementioned lines. Further follows a formal description of our approach.

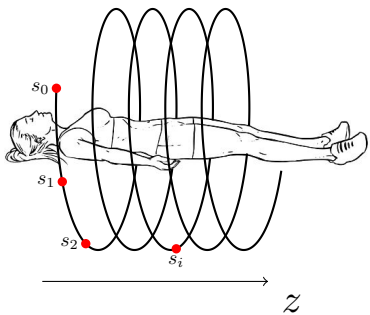
\begin{figure}[t]
  \centering
  \begin{tikzpicture}[scale=0.5, every node/.style={scale=1.5}]
    \put(0,12.5){
      \includegraphics[width=0.4\linewidth]{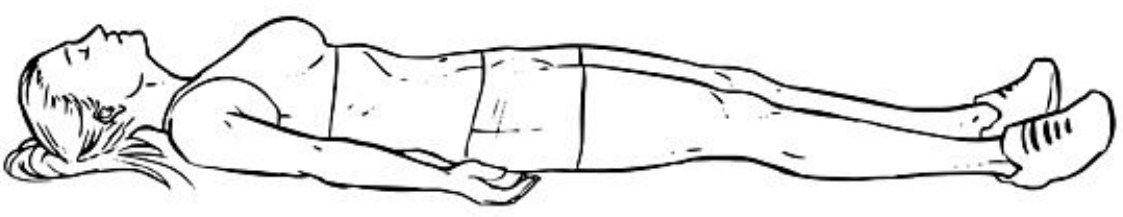}};
    \draw[->] (1,-2,0) -- (7,-2,0) node[below right,fill=none]{$z$};  
    \begin{axis}[
      hide axis,
      view={-20}{-20},
      zmax=80,
      xmax=2,
      ymax=2,
      height=12cm,
      clip=false
    ]   
    \begin{scope}[rotate=-45,transform shape]   
      \addplot3+[domain=0:1.015*pi,samples=500,
        samples y=0,black,no marks,ultra thick] 
        ({sin(deg(x))},{cos(deg(x))},{6*x/(pi)})
        node[name=A,red,circle,scale=0.5,fill,pos=0]{}
        node[name=B,red,circle,scale=0.5,fill,pos=0.25]{}
        node[name=C,red,circle,scale=0.5,fill,pos=0.5]{};
      \addplot3+[domain=1.133*pi:3.005*pi,samples=500,
        samples y=0,black,no marks,ultra thick] 
        ({sin(deg(x))},{cos(deg(x))},{6*x/(pi)});  
      \addplot3+[domain=3.14*pi:5.035*pi,samples=500,
        samples y=0,black,no marks,ultra thick] 
        ({sin(deg(x))},{cos(deg(x))},{6*x/(pi)})
        node[name=D,red,circle,scale=0.5,fill,pos=0.8]{};
      \addplot3+[domain=5.135*pi:7.0475*pi,samples=500,
        samples y=0,black,no marks,ultra thick] 
        ({sin(deg(x))},{cos(deg(x))},{6*x/(pi)});
      \addplot3+[domain=7.12*pi:9*pi,samples=500,
        samples y=0,black,no marks,ultra thick] 
        ({sin(deg(x))},{cos(deg(x))},{6*x/(pi)});
    \end{scope}
    \draw (A) node[anchor=east] {$s_0$};
    \draw (B) node[anchor=east] {$s_1$};
    \draw (C) node[anchor=east] {$s_2$};
    \draw (D) node[anchor=north] {$s_i$};
  \end{axis}
  \end{tikzpicture}    
  \caption{Source positions $s_i \in [0,2\pi)\times \Real$ are described by an angle $\varphi_i \in [0,2\pi)$ (very close to being uniform) and $z_i \in \Real$ offset along the rotation axis (not so uniform in practice).}
  \label{fig:helical_ct}
\end{figure}

\begin{figure}
    \centering
    \begin{tikzpicture}[scale=1, every node/.style={scale=1}]
    \draw[->] (-1,-2.2,0) -- (-1,2.2,0) node[left,fill=none]{$z$};
    \end{tikzpicture}
    \includegraphics[scale=0.4]{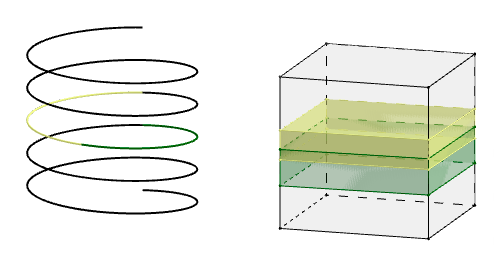}
    \caption{An illustration of splitting helical data (left) and image volume (right). One image sub-volume (a cube on the right) corresponds to the data obtained from the source positions on the helix marked with the same color (on the left). Note that image sub-volumes overlap, while data is partitioned without overlaps.
    }
    \label{fig:data_and_image_splitting}
\end{figure}
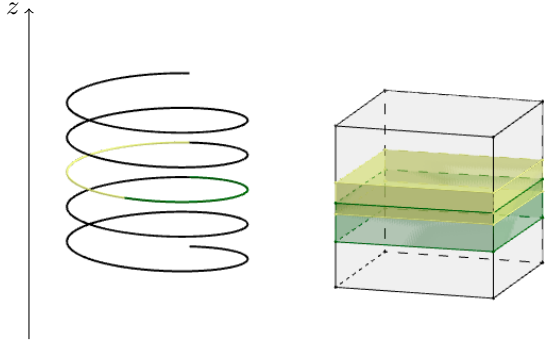

In practice, signals (images) and data measured with 3D helical \ac{CT} are digitized and represented by arrays.
The image space is $\RecSpace = \Real^{\ImageWidth \times \ImageHeight \times \NumSlices}$, where $\ImageWidth$ and $\ImageHeight$ are image width and height, respectively, and $\NumSlices$ is the number of slices along the rotation axis ($z$-axis). Note that $\NumSlices$ can be different for different scans, while the other two dimensions are fixed. The data space is $\DataSpace = \Real^{\NumSourcePos \times \DetWidth \times \DetHeight}$ where $\NumSourcePos$ is the total number of angles/source positions and $\DetWidth$ and $\DetHeight$ correspond to the number of detector columns and detector rows, respectively. As in the image space, the number of source positions along the trajectory ($\NumSourcePos$) can be different, while the dimensions of the detector are fixed.  \Cref{fig:helical_ct} provides a visualization of a helical trajectory of the radiation source around the patient.

We split the data space $\DataSpace$ into sub-spaces that correspond to the data acquired from the parts of a helical trajectory: 
\[ 
\DataSpace =  \DataSpace^1 \times \dots \times \DataSpace^j \times \dots \times \DataSpace^{\NumBlocks} \times \DataSpace^{\NumBlocks+1}. 
\]
Here, the $j$th data subspace is $\DataSpace^j := \Real^{\NumAngInLap \times \DetWidth \times \DetHeight}$, where number of angles in one part equals to $\NumAngInLap$ for $j = 1,2,\dots,\NumBlocks$. For simplicity fix $\NumAngInLap$ and discard the data corresponding to $\DataSpace^{\NumBlocks+1}$, i.e. the data observed from the source at the end of the trajectory. Thus, $\NumBlocks$ is the number of complete subspaces of the same dimension. We split the (truncated) data accordingly:
\[
\data = (\Lap^1, \Lap^2, \dots, \Lap^{\NumBlocks}).
\]
 
 Next, for each section of data, we consider a corresponding region in the image domain, from which this data has been obtained. Thus, from one image volume $\signal$ we extract $\NumBlocks$ overlapping sub-volumes
\[
(\Block^1, \Block^2, \dots, \Block^{\NumBlocks}) 
\quad \text{where $\Block^j \in \RecSpace^j$ for $j=1,\ldots,\NumBlocks$.}
\]
Each image subspace is $\RecSpace^j = \Real^{\ImageWidth \times \ImageHeight \times \ImageBlockThickness}$, where $\ImageBlockThickness$ is the minimal thickness of image sub-volume obtained by adding the physical detector height and the largest pitch.
 In this case, image subspace $\RecSpace^j$ contains all the line segments necessary to compute the data in subspace $\DataSpace^j$. Note that the image sub-volumes $\Block^j$ overlap, but the data sections $\Lap^j$ do not (see visualization in \cref{fig:data_and_image_splitting}).

We now define the restriction of the forward operator $\ForwardOp \colon \RecSpace \to \DataSpace$ to $\RecSpace^j$ as
\[
\ForwardOp^j: \RecSpace^j \mapsto \DataSpace^j,
\]
i.e., it is an intersection of rows and columns from $\ForwardOp$ corresponding to the $j$th data section and the $j$th sub-volume, respectively.
Next, we define projection operators 
\[
\ProjOp_{\RecSpace^j}: \RecSpace \mapsto \RecSpace^j, \quad s.t. \quad \ProjOp_{\RecSpace^j}(\signal) = \Block^j
\]
and 
\[
\ProjOp_{\DataSpace^j}: \DataSpace \mapsto \DataSpace^j, \quad s.t. \quad \ProjOp_{\DataSpace^j}(\data) = \Lap^j.
\]
Lastly, we define zero padding operators 
\[
\widetilde{\ProjOp}_{\RecSpace^j}: \RecSpace^j \mapsto \RecSpace, \quad
\text{and} \quad
\widetilde{\ProjOp}_{\DataSpace^j}: \DataSpace^j \mapsto \DataSpace.
\]
The proposed \ac{LPDh} method is now outlined in Algorithm~2. The networks $\DualMappingOriginal^i$ and $\PrimalMappingOriginal^i$ have the same architecture but different weights for $i=1, 2, \dots, M$. However, the same weights are used within the inner loop $j = 1,2, \dots, \NumBlocks$. This implies that the method can be applied with different numbers of data sections/line subsets $\NumBlocks$, which means that it can be applied to any length of the helical trajectory.

\begin{table}[t]
\centering
	\small
	\begin{tabular}[t]{@{\;}l}
	\hline
	\textbf{Algorithm 2} \acs{LPDh} \\
	\hline
	1: Choose initial primal and dual variables 
    \\ \quad
	$(\primal_0, \dual_0)=\texttt{init}(\sinogram)$, where $(\primal_0, \dual_0) \in (\PrimalSpace^\PrimalDim, \DataSpace^\DualDim)$ \\
	2: \textbf{For} $i=1,2,\dots,\NumSteps$ \textbf{do}: \\
	3: \hspace{0.2cm} \textbf{For} $j = 1, 2, \dots, \NumBlocks$ \textbf{do}: \\
        4: \hspace{0.7cm} $k = i\NumBlocks + j$ \\
        5: \hspace{0.7cm} Select section:
        $ \project{\dual}_{k-1} = \ProjOp_{\DataSpace^j}(\dual_{k-1}),\; \project{\data} = \ProjOp_{\DataSpace^j}(\data).$ \\
        6: \hspace{0.7cm} Select sub-volume:
        $ \project{\primal}_{k-1} = \ProjOp_{\RecSpace^j}(\primal_{k-1}).
        $\\[0.5em]
        7: \hspace{0.7cm} Dual:
    $ \dual_{k} = \dual_{k-1} + \widetilde{\ProjOp}_{\DataSpace^j}
    \left( 
      \DualMappingOriginal^{i} \left(
      \project{\dual}_{k-1}, 
      \ForwardOp^j \left(\project{\primal}^{(2)}_{k-1}\right), 
      \project{\data}
      \right)
    \right)$ \\[0.5em]
        8: \hspace{0.7cm} Primal: 
    $ \primal_{k} = \primal_{k-1} + \widetilde{\ProjOp}_{\RecSpace^j}
    \left(
      \PrimalMappingOriginal^i \left(
        \project{\primal}_{k-1},
        (\ForwardOp^{j})^{\ast}
         \left(\project{\dual}^{(1)}_{k-1} 
         \right)
      \right)
    \right)$ \\
	9: \textbf{return} $\primal_{\NumSteps \NumBlocks}^{(1)}$ \\[1mm]
    \hline
	\end{tabular}
\end{table}

The difference between \ac{LPD} and \ac{LPDh} is that the latter (partially) updates the primal variable after each partial update of the dual variable, while the former fully updates the dual variable and only then updates the primal variable. We empirically compare these two approaches in \cref{subsec:ablation}. We chose the section size to be equal to a single half-turn of a helical trajectory, so that back-propagation through one unrolled sub-iteration can be done on a \ac{GPU} with 24GB of memory. We use slightly smaller \acp{CNN} compared to \cite{Adler:2018aa}: The number of memory channels for the dual variable are $\DualDim=1$ instead of 5 and we use 16 channels instead of 32 in the hidden layers of the dual mapping $\DualMappingOriginal^i$. 

\subsection{Training and testing schemes} \label{sec:TrainTest}

We train the \ac{CNN}s $\DualMappingOriginal^i$ and $\PrimalMappingOriginal^i$ in \ac{LPD} and \ac{LPDh} using data corresponding to $\nlaps$ consecutive sections. More precisely, the loss for the training pair $(\signal_i,\data_i)$ is computed as
\begin{equation}\label{eq:loss_on_laps}
    \RecLoss\bigl(\RecOp_{\theta}(\ProjOp_{\DataSpace^{(q:q+\nlaps)}}(\data_i)),\ProjOp_{\RecSpace^{(q:q+\nlaps)}}(\signal_i) \bigr).
\end{equation}
where $\RecLoss$ is the standard $\ell_2$-norm and $q$ is a random number from $1$ till $\NumBlocks-\nlaps$. The projection operator defined in the data domain selects $\nlaps$ subsequent data sections, i.e. $\ProjOp_{\DataSpace^{(q:q+\nlaps)}} \colon \DataSpace \to \DataSpace^{q:q+\nlaps}$ where
\[
     \ProjOp_{\DataSpace^{(q:q+\nlaps)}}(\data) = (\Lap^q, \dots, \Lap^{q+\nlaps-1}) 
\]
The projection operator defined in the image domain selects the union of $\nlaps$ image sub-volumes corresponding to the data, i.e. $\ProjOp_{\RecSpace^{(q:q+\nlaps)}} \colon \RecSpace \to \RecSpace^{q:q+\nlaps}$ where
\[
     \ProjOp_{\RecSpace^{j}}\bigl(\ProjOp_{\RecSpace^{(q:q+\nlaps)}}(\signal)\bigr) = \Block^{q+j-1} 
     \quad\text{for $j =1,\dots,\nlaps$.}
\]

We hypothesise that this limited training is sufficient for the \ac{LPDh} to generalize to the case with more sections. This assumption is checked empirically in \cref{subsec:ablation}. 

An alternative strategy is to apply the method in a ``sliding window'' manner using the same number of sections as it was trained for. To combine the obtained reconstructions, we compute a weighted average, where a weight for each slice in a sub-volume depends on the distance of this slice to the center of the sub-volume. The further the slice is from the center, the smaller its weight is. More precisely, let us denote the center along the rotation axis of the reconstructed sub-volume with index $q$ by $z_c^q$ and its thickness by $z_t^q$. Then, the reconstruction $\estim{\signal}_z$ of the 2D slice $\signal_z$ is obtained from the independently reconstructed sub-volumes $\estim{\signal}_z^q = \RecOp_{\theta}\bigl(\ProjOp_{\DataSpace^{(q:q+\nlaps)}}(\data)\bigr)$ for $q = 1, \dots, \NumBlocks -\nlaps$ as follows:
\[
    \estim{\signal}_z = \frac{\sum_{q=1}^{\NumBlocks} w_{z,q} \estim{\signal}_z^q}{\sum_{q=1}^{\NumBlocks} w_{z,q}}
\]
where
\[
    w_{z,q} = 
    \begin{cases}
      1 - \dfrac{2}{z_t^q } |z-z_c^q|,
      &\text{ if $|z-z_c^q| \leq z_t^q / 2$} \\[1em]
      0 &\text{ otherwise.}
    \end{cases}
\]

\section{Experiments and results}

First, we use simplified training data to select hyper-parameters and to compare the different approaches described in the previous section in a so-called ablation study. Then, we train the \ac{LPDh} using realistic training data and perform quantitative comparison to a baseline method. Finally, we evaluate the trained network on the (pre-processed) real clinical data.

\subsection{Data}\label{subsec:data}

We simulate experimental data using \ac{LDCT} data set \cite{mccollough2020} provided by the Cancer Imaging Archive. Furthermore, we evaluate the proposed methods on the \emph{real} projection data from this data set. The data set contains helical \ac{CT} scans of 299 patients involving head, thorax, and abdomen acquired using scanners from Siemens Healthcare and GE Healthcare. Due to differences in scanning protocols for different vendors and different body parts, we chose to focus only on scans of the abdomen (45~in total) acquired using a Siemens system. Of these, 40 are used for training, one for validation (`L134'), and four for testing (`L072', `L019', `L116', `L150'). For each case, one has full dose and reduced dose (25\%) projection data along with corresponding reference 3D reconstructions.

\paragraph{Simulated data for training}
The 3D reconstructions obtained from full dose \ac{CT} in the above mentioned data set were used as digital phantoms to generate low-dose helical \ac{CT} data in the following way. 
First, the input 3D reconstruction $\signal$, which is given in the Hounsfield units, is re-scaled as 
\begin{equation*}
    \signal = \Bigl(\frac{\signal_{\text{HU}}}{1000} + 1\Bigr)  \mu_0,
\end{equation*}
where $\mu_0 := 0.0192$~mm${}^{-1}$ is the X-ray attenuation of water at the mean X-ray energy of 70~keV. Next, noise-free tomographic data $\sinogram = \ForwardOp(\primal)$ is simulated using the ray transform with the original acquisition geometry provided together with the data. Next, the data is corrupted by Poisson noise 
\begin{equation*}
    H_{\text{noisy}} = \operatorname{Poisson}\Bigl(H_0 e^{-\ForwardOp(\primal)}\Bigr),
\end{equation*}
where $H_0$ is the number of photons per pixel available with the data. 
Finally, the data is linearized so that $\sinogram \approx \ForwardOp(\primal)$ where
\begin{equation*}
\sinogram = - \ln\left(\frac{1}{H_0} H_{\text{noisy}}\right).
\end{equation*}

 The full dose reconstructions included in the data set (our ground truth) are slightly truncated from both ends along the rotation axis comparing to the region that was illuminated to generate the projection data. This is done, because reconstruction in those region is only partial, due to the lack of data from all angles. Therefore, when we simulate the data, we discard source positions in the beginning and at the end of a scan, leaving only those source positions, whose rays cross the ground truth volume and do not cross missing regions.   

\paragraph{Simplified geometries for the ablation study} In order to perform the ablation experiments in a shorter time frame, we simplify the original acquisition geometries included in the data set by using a flat detector and not using \ac{FFS} (we disregard every second angle). 

\paragraph{Real data} The simulation process described above is not perfect (even with real acquisition geometries and photon statistics).
Hence, in order to apply the reconstruction method trained on simulated data, we need to apply corrections to the real data.
First of all, the \ac{LDCT} reference reconstructions, that we use as phantoms to generate the training data, do not cover objects that are outside the scanner's field of view, such as cables, blankets, and the table.
Therefore, our method is not trained to reconstruct those objects.
Secondly, visual examination of the real data shows that for some source angles there appears to be an additive error, which is almost constant along the first detector axis (detector width). This likely comes from normalizing data with incorrect gain estimates \cite{thibault2007correction}.

To address both of these issues we seek our own reference reconstruction using the full dose data.  
Since we don't have access to other methods we construct a new reference reconstruction by solving the optimization problem:
\begin{equation}\label{eq:new_ref}
    \min_{\primal,\fluxCorr} \DataLoss(\ForwardOp(\primal), \data + \fluxCorr), 
\end{equation}
where $\fluxCorr \in \Real^{\NumSourcePos \times \DetHeight}$ is a gain correction for each source position and detector row and the data discrepancy $\DataLoss \colon \DataSpace \times \DataSpace \to \Real$ is a weighted $l_2-$norm \cite{elbakri2002statistical}:
\begin{equation}\label{eq:dataLoss}
    \DataLoss\bigl(\ForwardOp (\signal), \data\bigr) = \sum_{i = 1}^{|\DataSpace|} w_i \bigl\Vert \ForwardOp (\signal)_i - \data_i \bigr\Vert_2^2
    \quad\text{where $w_i := e^{-\data_i}$.} 
\end{equation}
We solve \cref{eq:new_ref} using Nesterov's accelerated gradient descent \cite{nesterov1983method} for 200 iterations. We use the obtained reconstruction to correct the real data as follows
\begin{equation}\label{eq:dataCorr}
    \data_{corr} = \data_{real} - \ForwardOp(\mathcal{M}(\estim{\primal})) + \estim{\fluxCorr}
\end{equation}
where  $\mathcal{M} \colon \RecSpace \to \RecSpace$ is a an operator that masks the central part of the reconstruction, so that only the truncated regions are projected, and $\estim{\primal}, \estim{\fluxCorr}$ are obtained by minimizing \cref{eq:new_ref} as described above.

We hypothesize that the corrections \cref{eq:dataCorr} could be (as well) performed using a full (not truncated) weighted \ac{FBP} reconstruction for $\estim{\primal}$ and $\data - \ForwardOp(\estim{\primal})$ averaged along the first detector axis for $\estim{\fluxCorr}$. Alternatively, the training could be done on full (not-truncated) phantoms.

Lastly, the measured data has much higher resolution in the dimension along the rotation axis than the reference reconstructions. Minimizing data discrepancy  \cref{eq:new_ref} for image volumes $\signal \in  \RecSpace$ that are too coarse results in typical ``undershooting'' artifacts near structures with high attenuation such as bones. To avoid this, we smooth the data along this dimension by applying a convolution with a normalized triangle filter with half width equal twice the size of reconstructed voxels.

\subsection{Implementation details}\label{subsec:implementation_details}

The proposed method is implemented in Python with specific C++/CUDA based libraries for computationally demanding tasks. 
An example of the latter is \ac{CT}-related components, which are implemented in the \ac{ODL} \cite{adlerODL} with ASTRA \cite{van2016fast} as back-end for computing forward/back-projections. The neural network layers and training are implemented using PyTorch \cite{paszke2017automatic}.

The neural networks used in \ac{LPD} and \ac{LPDh} are almost identical to the original method \cite{Adler:2018aa} regarding structure and hyper-parameters. A three-layer \ac{CNN} is applied in both image and projection domain at each unrolled iteration, i.e., $\PrimalMappingOriginal^i$ and $\DualMappingOriginal^i$ have two hidden convolutional layers with $32$ and $16$ filters, respectively, and one convolutional output layer. \acp{ReLU} are used as activation functions after hidden layers. Training is performed with the Adam optimizer \cite{Kingma:2014aa} using the $\ell_2$-distance to the ground-truth described in eq.~\eqref{eq:loss_on_laps}. The initial learning rate is set to $5\cdot 10^{-4}$ with cosine annealing. Since the training samples become larger with the increased number of sections, we define the total number of training iterations to be $2\cdot 10^5 / (10 + 5 \cdot N_s)$, where $10 + 5\cdot N_s$ is a rough estimate of the number of slices in a target reconstruction for a given number of sections $N_s$. However, we use $10^4$ iterations, when training with the real geometries.

Theoretically, the size of image sub-spaces $\RecSpace^j$ depends on the pitch. In practice, the pitch can vary across different scans and also within each scan (we observe values within the range 17.18~mm to 34.5~mm in our training and testing set). The method must therefore be robust w.r.t.\@ these variations, assuming they are not too big. For simplicity, we set $\ImageBlockThickness = 16$, which is a suitable value for most of the scans. 

Quantitative performance of all methods is measured by calculating the \ac{PSNR} and the \ac{SSIM} between the ground truth image and its corresponding reconstruction. However, since the data is truncated at the beginning and the end of acquisition geometry, we discard first 8 and last 8 reconstructed 2D slices, since reconstruction in those regions is only partial. 

\paragraph{ Baseline method} We compare our approach to a variational model with Huber regularizer, i.e., reconstruction obtained from $\RecOp_{\theta} \colon \DataSpace \to \RecSpace$ that is defined as 
\begin{equation}\label{eq:VarReg}
  \RecOp_{\theta}(\data) \in \argmin_{\signal} \Bigl\{
    \DataLoss\bigl(\ForwardOp(\signal),\data \bigr) +  \lambda \RegFunc_{\theta}(\signal)
  \Bigr\}.
\end{equation}
Here, the data discrepancy is given as in \cref{eq:dataLoss} and the regularizer $\RegFunc_{\theta} \colon \RecSpace \to \Real$ is smooth relaxation of the non-smooth $\ell_1$-norm that makes up the \ac{TV} regularizer \cite{huber2004robust}:
\[
\RegFunc_{\theta}(\signal) := \sum_{i=1}^{2\vert \RecSpace \vert} h_{\theta}\bigl( \vert \nabla\signal_i \vert \bigr)
\quad\text{where }
h_{\theta}(t) := 
\begin{cases}
  t^2/(2\theta),  & t \le \theta \\
  t - \theta/2, & t > \theta. 
  \end{cases}
\]
In our experiments, we set the regularization parameter and the Huber parameter to $\lambda = 0.15$ and $\theta = 0.0012$, respectively. 
The optimization in \cref{eq:VarReg} is solved using Nesterov's accelerated gradient descent \cite{nesterov1983method} that is stopped (early) after 200 iterations. We initialize the optimization with an approximate \ac{FBP} algorithm implemented in \ac{ODL}, which does not account for \ac{FFS}. We chose the hyper-parameters and the number of iterations that lead to optimal performance on the validation data. In particular, early stopping does not only improve upon computational feasibility, it has also been shown to provide an additional regularising effect as outlined in \cite{effland2020variational}. 

\subsection{Ablation study}\label{subsec:ablation}

In this section we empirically compare the methods described in \cref{sec:method}. Since training of the neural networks on a single high-end consumer \ac{GPU} takes weeks, we use data simulated using simplified geometries (as described in \cref{subsec:data}). In \cref{tab:abl} we evaluate the performance of \ac{LPD} and \ac{LPDh} trained with different number of sections. 

First of all, we can see that \ac{LPDh} unambiguously outperforms \ac{LPD} without increase in total computational time for inference. Secondly, both \ac{LPD} and \ac{LPDh} perform poorly, if trained only using 2 subsequent sections. In this cases, it is much more beneficial to apply the method in a sliding window manner. However, as the number of sections during the training increases, the generalization to an unlimited number of sections (25--72 in our test cases) improves greatly. If 6 sections are used, there is almost no difference in performance between the sliding window testing strategy and the regular forward pass, suggesting that almost perfect generalization is achieved in this case. Ultimately, we can see that \ac{LPDh} with 6 sections provides very good performance without a relevant increase in estimation time. Still, the best performing approach is \ac{LPDh} trained with 4 sections and tested in sliding window manner. We evaluate the latter in the next sections using realistic simulated data and real data.

\begin{table}[thb]\label{tab:abl}
    \centering
    \begin{tabular}{l r r r}
         &  \acs{PSNR} & \acs{SSIM} & time, min\\
    \toprule
    \acs{LPD}, 2 sections & 28.74 & 0.524 & 9.9\\
    \acs{LPD}, 2 sections, sliding window & 39.61  & 0.964& 23.5\\
    \acs{LPDh}, 2 sections & 24.94 &  0.477 & 9.9\\
    \acs{LPDh}, 2 sections, sliding window & 40.88  & 0.972 & 19.1\\
    \hline
    \acs{LPD}, 4 sections & 38.83 &0.957 & 9.9\\
    \acs{LPD}, 4 sections, sliding window & 39.56 & 0.964 & 44.1\\
    \acs{LPDh}, 4 sections & 40.59 & 0.970 & 9.7 \\
    \textbf{\acs{LPDh}, 4 sections, sliding window} & 41.37 & 0.975 & 35.9\\
    \hline
    \acs{LPD}, 6 sections & 39.52 & 0.962 & 9.7\\
    \acs{LPD}, 6 sections, sliding window & 39.86 & 0.966 & 64.2\\
    \acs{LPDh}, 6 sections & 40.95 & 0.972 & 10.2\\
    \acs{LPDh}, 6 sections, sliding window & 41.02 & 0.974 & 52.4\\
    \bottomrule
    \end{tabular}
    \caption{Performance metrics for \ac{LPD} and \ac{LPDh} using different number of sections during training. Averages over four test patients are reported.}
\end{table}

\subsection{Results}\label{subsec:results}

We here evaluate \ac{LPDh} with 4 sections, since it showed the best performance in \cref{subsec:ablation}. It is trained against simulated data with geometries and noise levels that correspond to those in the \ac{LDCT} data set. \Cref{tab:results} lists quantitative performance results in terms of \ac{PSNR} and \ac{SSIM} as well as average execution time. \Ac{LPDh} clearly  outperforms the baseline Huber regularization, while being an order of magnitude faster.

\begin{table}[thb]
    \centering
    \begin{tabular}{l r r r}
         & \acs{PSNR}  & \acs{SSIM} & time, min\\
    \toprule  
    Huber reg. & 44.65 & 0.981 & 720 \\
    \ac{LPDh} & 45.80 & 0.986 & 40 \\
    \ac{LPDh}, sliding window & 46.19 & 0.987 & 127 \\
    \bottomrule
    \end{tabular}
    \caption{Performance metrics for different reconstruction methods on realistic data. Averages over four test patients are reported.}
    \label{tab:results}
\end{table}

Next, we perform visual comparison of the proposed methods applied to the real data (corrected as described in \cref{subsec:data}) in \cref{fig:TransverseCS}. We present two baseline reconstructions: the low dose reference included in the data set and the iterative reconstruction with Huber regularization. We also show two full dose reconstructions: one that is included in the data set and our own reference obtained from the full dose data by addressing problem \cref{eq:new_ref}. 

\section{Limitations}\label{sec:discussion}

We do not present a quantitative comparison of reconstructions obtained from the real data.
The reason is that the \ac{LDCT} full dose reference does not represent the actual ``ground truth'' with sufficient accuracy.
\Cref{fig:TransverseCS} shows that the iterative reconstruction computed in a time-unconstrained setting seems to provide a superior recovery of small details. However, it does not mean that the quality of our reference reconstruction cannot be challenged.
It is, e.g., likely to contain more noise than the \ac{LDCT} full dose reference. We are therefore hesitant to use it as the ground truth.

Yet another issue that undermines quantitative comparison is that full dose and low dose reconstructions obtained with the same method will have additional similarities that result from implicit/explicit prior information imposed by the method, let's call it method's bias. Thus, the low dose \ac{FBP} reconstruction is closer in \ac{PSNR} to the full dose \ac{FBP} reconstruction, while iterative reconstructions, Huber regularization and \ac{LPDh}, are closer to the iterative full dose reconstruction (our full dose reference).

Furthermore, the issues described above should be accounted for when training the method in a supervised manner with the real data. We tried to train \ac{LPDh} using the \ac{LDCT} full dose reference, but the performance was similar (slightly worse) to U-Net post-processing of the low dose reference. Nevertheless, we have seen in figure \cref{fig:TransverseCS} that the full dose reference does not contain all the details that \ac{LPDh} is able to reconstruct after being trained on \emph{simulated} data. Moreover, quantitative comparison confirms that \ac{LPDh} reconstructions are closer to our own reference than both \ac{LDCT} reference reconstructions. One option would be to train \ac{LPDh} using full dose iterative reconstructions as proxies for ground truth. However, obtaining such reconstructions for the whole data set requires a lot of computational resources and the quality of reconstruction would still be questioned. 

To summarize, it is clear that there are differences between full dose reconstructions obtained with different methods. Furthermore, none of the methods can perfectly recover the actual ``ground truth''. 
Obtaining a good proxy for the ground truth using full dose data in a time-unconstrained setting requires a separate investigation, which should be ideally performed by the maintainers of \ac{LDCT} data set. An alternative approach is to avoid defining a ground truth and consider quality metrics based on the data discrepancy in the projection domain. 
This was done in \cite{kosomaa2022projection} where some randomly selected projections are left out during the reconstruction process, so that they can be used to compute the unsupervised data discrepancy loss.

The final limitation is the lack of freely available software suitable for reconstructing the clinical \ac{CT} data, in particular the ones from the \ac{LDCT} challenge. For instance, the open source software library \emph{FreeCT} \cite{hoffman2016freect_wfbp} that implements the \ac{wFBP} algorithm from \cite{stierstorfer2004weighted} currently does not support \ac{FFS}. As a consequence, we used iterative reconstruction with Huber regularizer as the baseline method, even though the method is not practical due to its execution time. 
Like all iterative reconstruction algorithms, \ac{LPDh} is computationally much more demanding than analytic methods and thus inherently slower. However, keeping in mind that iterative methods are clinically available and that the learned components account only for $12\%$ of our total execution time, this aspect should not prevent the method from clinical use.
Finally, it is possible that with a good initialization of \ac{LPDh} method (instead of zero-initialization) the number of unrolled iterations could be reduced. Such initialization could be applied if high quality analytic methods, such as the ones used by vendors, were available.

\newcommand\x{1.0} 
\newcommand{\spyonimtwo}{  
  \spy on (0.0*\x,-0.55*\x) in node at (1.4*\x,-1.4*\x);
  \spy on (-1.0*\x,0.6*\x) in node at (-1.4*\x,-1.4*\x);
  \spy on (0.9*\x,0.2*\x) in node at (1.4*\x,1.4*\x);
  \begin{pgfonlayer}{foreground}
     \node [draw, red, thick, fill=white] at (1.82*\x,-0.7*\x) {C};
     \node [draw, red, thick, fill=white] at (-1.82*\x,-0.7*\x) {B};
     \node [draw, red, thick, fill=white] at (1.82*\x,0.7*\x) {A};
     \draw [-stealth, red, thick](1.4*\x,-1.4*\x) -- (1.45*\x,-1.6*\x);
     \draw [-stealth, red, thick](-1.4*\x,-1.4*\x) -- (-1.24*\x,-1.25*\x);
  \end{pgfonlayer}{foreground}}

\begin{figure*}
\centering
\begin{minipage}{.32\textwidth}
\centering
\begin{tikzpicture}[spy using outlines={
   rectangle, 
   red, 
   magnification=2.5,
   size=0.3\linewidth, 
   connect spies}]
  \node{\includegraphics[width=\linewidth, angle=180]
    {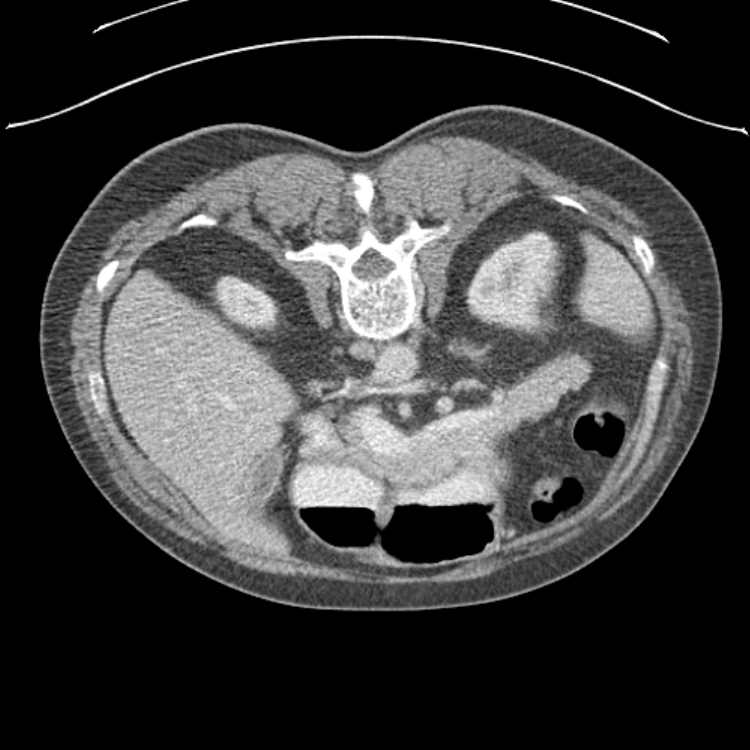}};
    \spyonimtwo
\end{tikzpicture}
\\ Low dose reference
\end{minipage}
\hfill
\begin{minipage}{.32\textwidth}
\centering
\begin{tikzpicture}[spy using outlines={
   rectangle, 
   red, 
   magnification=2.5,
   size=0.3\linewidth, 
   connect spies}]
  \node{\includegraphics[width=\linewidth, angle=180]
    {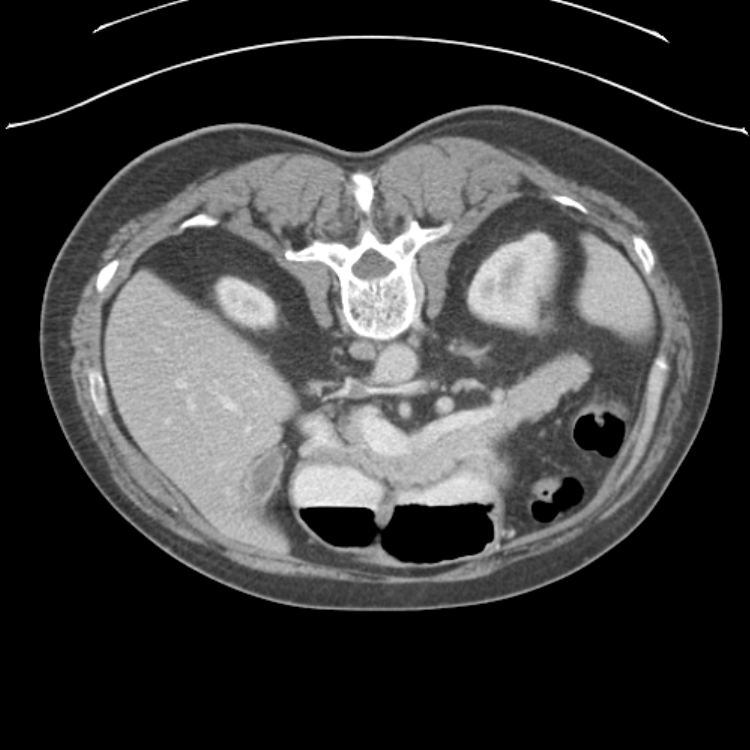}};
  \spyonimtwo
\end{tikzpicture}
\\ Full dose reference
\end{minipage}
\hfill
\begin{minipage}{.32\textwidth}
\centering
\begin{tikzpicture}[spy using outlines={
   rectangle, 
   red, 
   magnification=2.5,
   size=0.3\linewidth, 
   connect spies}]
  \node{\includegraphics[width=\linewidth, angle=180]
    {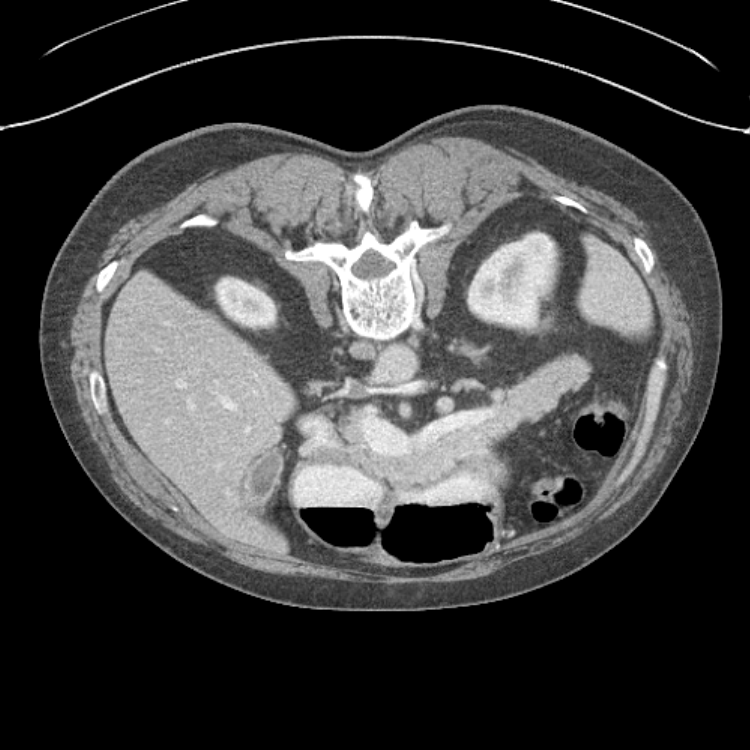}};
  \spyonimtwo
\end{tikzpicture}
\\ Our full dose reference
\end{minipage}
\par\medskip
\begin{minipage}{.32\textwidth}
\centering
\begin{tikzpicture}[spy using outlines={
   rectangle, 
   red, 
   magnification=2.5,
   size=0.3\linewidth, 
   connect spies}]
  \node{\includegraphics[width=\linewidth, angle=180]
    {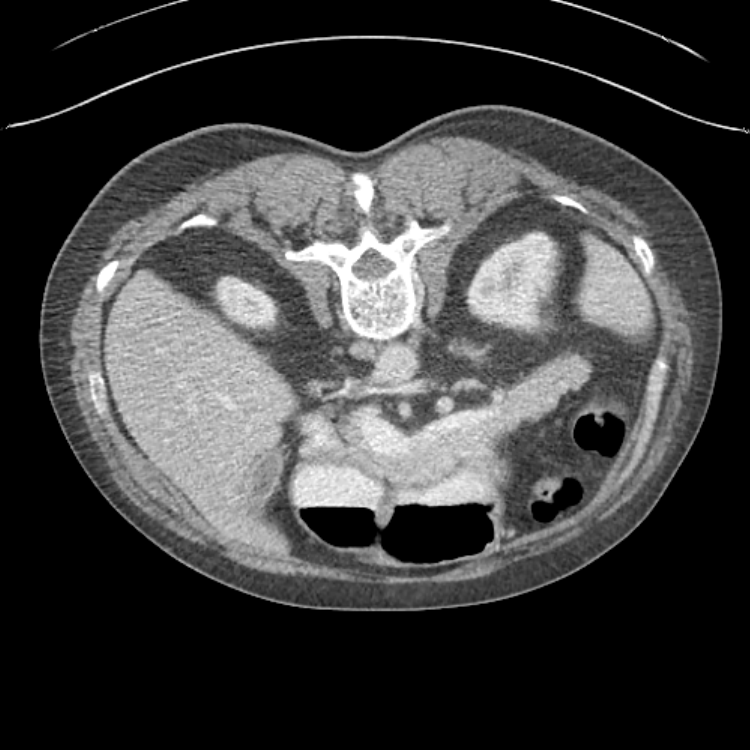}};
  \spyonimtwo
\end{tikzpicture}
\\ Huber regularization
\end{minipage}
\hfill
\begin{minipage}{.32\textwidth}
\centering
\begin{tikzpicture}[spy using outlines={
   rectangle, 
   red, 
   magnification=2.5,
   size=0.3\linewidth, 
   connect spies}]
  \node{\includegraphics[width=\linewidth, angle=180]
    {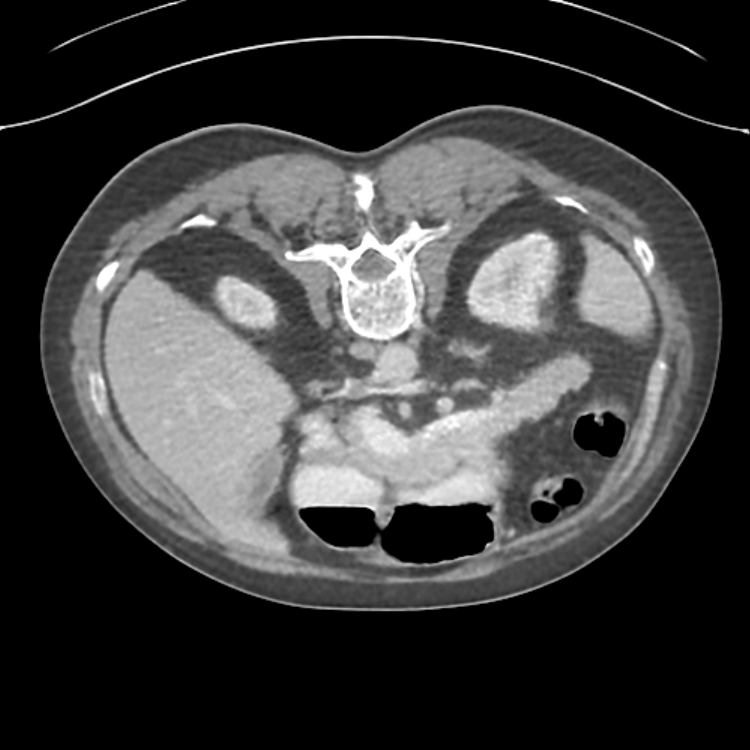}};
  \spyonimtwo
\end{tikzpicture}
\\ \acs{LPDh}
\end{minipage}
\hfill
\begin{minipage}{.32\textwidth}
\centering
\begin{tikzpicture}[spy using outlines={
   rectangle, 
   red, 
   magnification=2.5,
   size=0.3\linewidth, 
   connect spies}]
  \node{\includegraphics[width=\linewidth, angle=180]
    {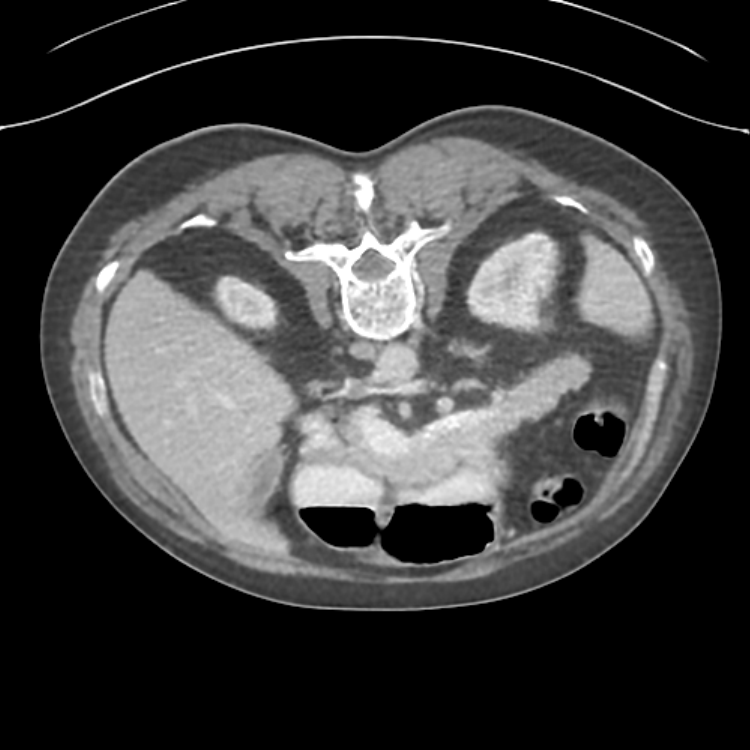}};
  \spyonimtwo
\end{tikzpicture}
\\ \ac{LPDh}, sliding window
\end{minipage}
\caption{Transverse cross section of 3D reconstructions obtained from different methods (level 0 HU, window 400 HU). Full dose reference and our reference are obtained from full dose data and serve as proxies for ``ground truth''. Region (A) shows improved denoising with \ac{LPDh} methods. Region (B)  shows reconstruction of small high contrast structures. We can see that \ac{LPDh} shows the connective tissue in contrast to the full dose reference included in the data set. Examination of our own full dose reference (obtained through iterative reconstruction) reveals that this is not a mistake. Region (C) shows that \ac{LPDh} methods still suffer from undershooting artifacts, which were almost (but not fully) removed by smoothing the data as described in \cref{subsec:data}. }
\label{fig:TransverseCS}
\end{figure*}

\newcommand{\figsize}{0.3}
\newcommand{\scale}{\figsize/0.24*\x}
\newcommand{\spyonimcor}{  
  \spy on (-1.0*\scale,1.6*\scale) in node at (0.8*\scale,2.3*\scale);
  \spy on (0.4*\scale,0.6*\scale) in node at (0.8*\scale,-2.3*\scale);
}
\newcommand{\spyonimsag}{  
  \spy on (0.35*\scale,1.24*\scale) in node at (0.8*\scale,2.3*\scale);
  \spy on (0.5*\scale,0.55*\scale) in node at (0.8*\scale,-2.3*\scale);
}
\begin{figure*}
\centering
\begin{minipage}{\figsize\textwidth}
\centering
\begin{tikzpicture}[spy using outlines={
   rectangle, 
   red, 
   magnification=2.5,
   size=0.4\linewidth, 
   connect spies}]
\node{\includegraphics[width=\linewidth]{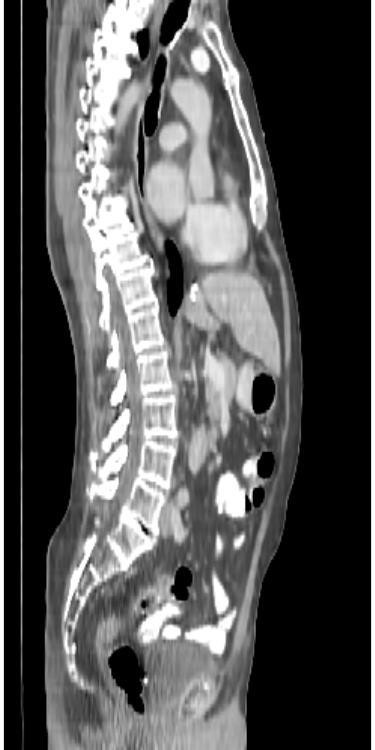}};
\spyonimcor
\end{tikzpicture}
\\ \ac{LPDh}
\end{minipage}
\hfill
\begin{minipage}{\figsize\textwidth}
\centering
\begin{tikzpicture}[spy using outlines={
   rectangle, 
   red, 
   magnification=2.5,
   size=0.4\linewidth, 
   connect spies}]\node{\includegraphics[width=\linewidth]{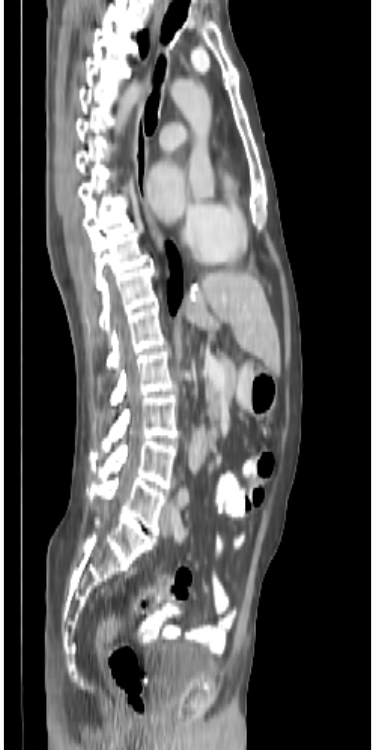}};
\spyonimcor
\end{tikzpicture}
\\ \ac{LPDh}, sliding win.
\end{minipage}
\hfill
\begin{minipage}{\figsize\textwidth}
\centering
\begin{tikzpicture}[spy using outlines={
   rectangle, 
   red, 
   magnification=2.5,
   size=0.4\linewidth, 
   connect spies}]
\node{\includegraphics[width=\linewidth]{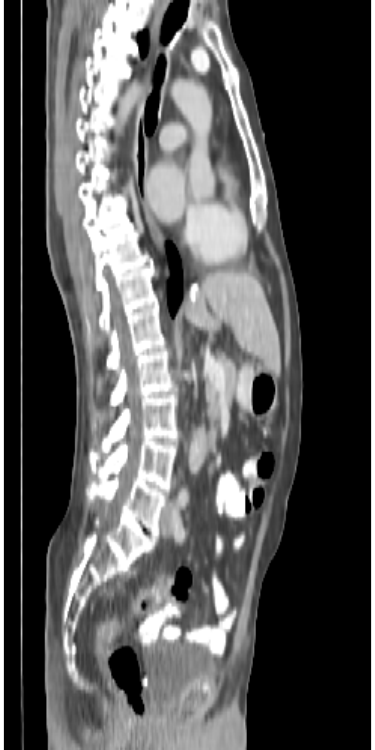}};
\spyonimcor
\end{tikzpicture}
\\ \ac{LPDh}, simulated
\end{minipage}
\par
\begin{minipage}{\figsize\textwidth}
\centering
\begin{tikzpicture}[spy using outlines={
   rectangle, 
   red, 
   magnification=2.5,
   size=0.4\linewidth, 
   connect spies}]
\node{\includegraphics[width=\linewidth]{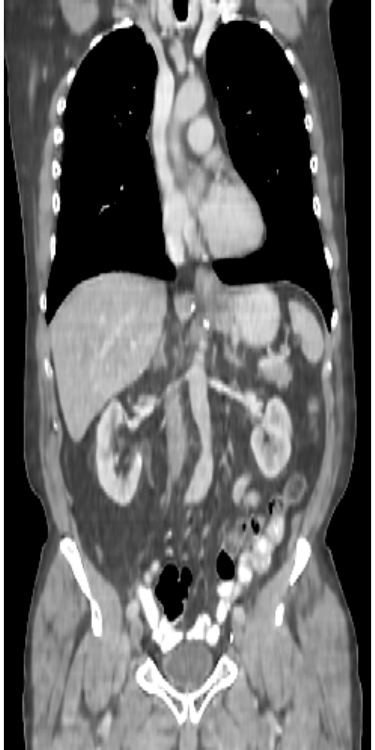}};
\spyonimsag
\end{tikzpicture}
\\ \ac{LPDh}
\end{minipage}
\hfill
\begin{minipage}{\figsize\textwidth}
\centering
\begin{tikzpicture}[spy using outlines={
   rectangle, 
   red, 
   magnification=2.5,
   size=0.4\linewidth, 
   connect spies}]
\node{\includegraphics[width=\linewidth]{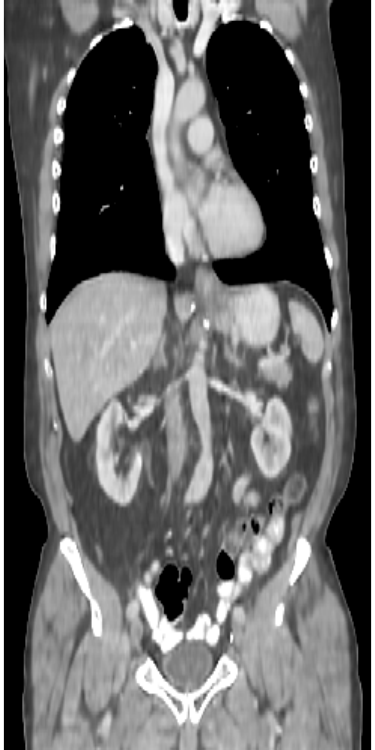}};
\spyonimsag
\end{tikzpicture}
\\ \ac{LPDh}, sliding win.
\end{minipage}
\hfill
\begin{minipage}{\figsize\textwidth}
\centering
\begin{tikzpicture}[spy using outlines={
   rectangle, 
   red, 
   magnification=2.5,
   size=0.4\linewidth, 
   connect spies}]
\node{\includegraphics[width=\linewidth]{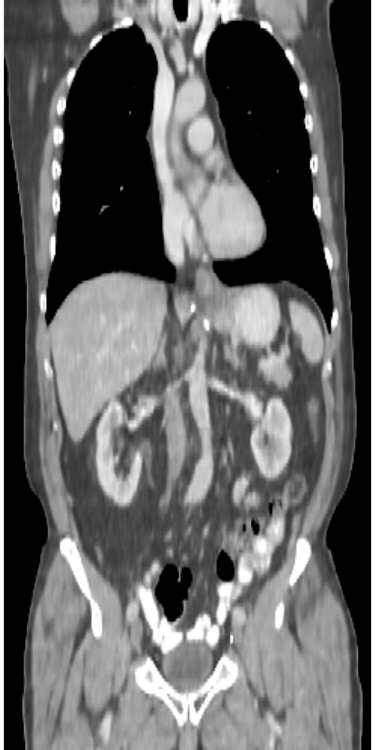}};
\spyonimsag
\end{tikzpicture}
\\ \ac{LPDh}, simulated
\end{minipage}
\caption{Sagittal and coronal cross sections of 3D reconstructions obtained with \ac{LPDh} (level 0 HU, window 400 HU) from the real and simulated data. Splitting of the image volume during the reconstruction leads to subtle horizontal streak artifacts, which are less noticeable in the reconstruction obtained in a sliding window manner. The remaining small differences between simulated and the real data seems to be partially responsible for the appearance of the artifacts, since the artifacts are almost invisible in the reconstructions from the simulated data. }
\label{fig:LongitudalCS}
\end{figure*}

\section{Conclusions}
We have proposed \ac{LPDh} -- a new algorithm in the family of Learned Primal-Dual methods that is suited for the reconstruction of helical \ac{CT} data. 
We overcome the main challenge -- large \ac{GPU} memory requirements of the original \ac{LPD} method, and train \ac{LPDh} on a \emph{single consumer \ac{GPU}}.

To achieve this, we address the problem on several levels. First is to apply gradient check-pointing to unrolled iterations. Second is to split geometry and data into sections, so that an unrolled iteration w.r.t.\@ one section fits the \ac{GPU} memory. Finally, we train the method on smaller samples of the data that correspond to a few helical turns. Although our method generalizes very well to the data that is an order of magnitude larger than the training samples, we find that applying the method in a sliding window manner can additionally boost the performance at the cost of computational time. 

Training and quantitative performance evaluation has been done on data from realistic simulations. Qualitative performance has been assessed on clinical data from the \ac{LDCT} data set.

\section{Acknowledgements}

The computations were partially enabled by resources provided by the National Academic Infrastructure for Supercomputing in Sweden (NAISS) at Chalmers partially funded by the Swedish Research Council through grant agreement no. 2022-06725.

\bibliographystyle{plain}
\bibliography{references}

\end{document}